\documentclass[aps,prl,onecolumn,groupedaddress]{revtex4}
\usepackage{amsmath}
\usepackage{amssymb}
\usepackage{graphicx}
\usepackage{bm}

\begin{document}

\title{Symmetry-restoring quantum phase transition in a two-dimensional spinor condensate.}

\author{A. L. Chudnovskiy}
\affiliation{1. Institut f\"ur Theoretische Physik, Universit\"at Hamburg,
Jungiusstr 9, D-20355 Hamburg, Germany}

\author{V. Cheianov}
\affiliation{Instituut-Lorentz, Universiteit Leiden, P.O. Box 9506, 2300 RA Leiden, The Netherlands}

\date{\today}

\begin{abstract}
Bose Einstein condensates of spin-1 atoms are known to exist in two different phases, both having spontaneously broken spin-rotation symmetry, a ferromagnetic and a polar condensate.  Here we show that in two spatial dimensions it is possible to achieve a quantum phase transition from a polar condensate into a singlet phase symmetric under rotations in spin space. This can be done by using particle density as a tuning parameter. Starting from the polar phase at high density the system can be tuned into a strong-coupling 
intermediate-density point where the phase transition into a symmetric phase takes place. 
By further reducing the particle density the symmetric phase can be continuously deformed into a Bose-Einstein condensate of singlet atomic pairs. We calculate the region of the parameter space where such a molecular phase is stable against collapse.
\end{abstract}


\maketitle

Spinor atomic quantum gases provide a platform for experimental realization of wide variety of quantum phases with exotic properties. The high degree of control over the inter-atomic interactions allows to investigate the magnetic quantum ground states that often have no analogies in solid state systems \cite{Stamper-KurnReview2013,Lewenstein07,Hermele09,Hermele10,Hermele16,Demler02,Wen02}. Exotic properties of quantum ground state in the Bose gases are related to a nontrivial structure of internal degrees of freedom of the Bose condensate wave function. The simplest nontrivial example of the rich magnetic phase diagram constitutes the gas of spin-1 bosons. The Hamiltonian of this system is given by 
\begin{equation}
H= \int d\mathbf x \left[ \frac{\hbar^2}{2m} \nabla \psi_a^\dagger \nabla \psi_a + 
\frac{c_0}{2} \psi_a^\dagger \psi_b^\dagger \psi_b \psi_a + 
\frac{c_2}{2} \psi_a^\dagger \psi_{a'}^\dagger \psi_{b'} \psi_{b} \mathbf F_{ab} \cdot \mathbf  F_{a' b'}\right],
\label{Hamiltonian}
\end{equation}
where the spin-index $a$ assumes the values $1, 0, -1$, and  the matrix-valued vector $\mathbf F$ consists of the $S=1$ representation of $SU(2)$ group generators. The interaction constants $c_0$ and $c_2$ describe the scalar and spin interactions respectively. Since the interaction in Eq. (\ref{Hamiltonian}) is spin-conserving, it is convenient to represent the interaction term in Eq. (\ref{Hamiltonian}) according to the spin-0 and spin-2 scattering channels (the s-wave scattering in the spin-1 channel is suppressed by the symmetry of the bosonic wave function). Thereby the effective interaction constants become $g_0=c_0-2c_2$, and $g_2=c_0+c_2$. 
Mean field analysis of the Hamiltonian (\ref{Hamiltonian}) reveals two stable zero-temperature phases \cite{Ho98}: the ferromagnetic BEC condensate for $c_2<0$,  and the polar condensate for anti-ferromagnetic interactions, $c_2>0$. In both phases, the $SU(2)$ symmetry is spontaneously broken, resulting in divergent magnetic susceptibilities at the transition, and gapless spin-waves as elementary low-energy excitations. 

Curiously, although there seem to be no fundamental reasons precluding the existence of the $SU(2)$-symmetric Bose-condensate,  such a singlet phase having no gapless spin-waves is not present on the mean field phase diagram. An attempt to construct a singlet phase by taking into account quantum dynamics of condensate's zero modes was made in Ref. \cite{Bigelow98}. It was shown that such quantum effects may favor a spin-singlet vacuum state. However, the resulting vacuum is unstable against an infinitesimal  magnetic field in the thermodynamic limit \cite{Ho-Yip99}, signifying a spontaneously broken symmetry \cite{Bogoliubov_Book}. 
Therefore, the mechanism discussed in Ref. \cite{Bigelow98} does not lead to the formation of an $SU(2)$-symmetric phase.


In this paper we show that a quantum phase transition from the polar condensate to an $SU(2)$-symmetric phase can be achieved in two dimensions by varying the concentration of atoms. Our conlcusion is based on the interpolation between two limiting cases, each  admitting for controllable analytical treatment. The limit of high particle density is amenable to  the mean field approach of  Ref. \cite{Ho98}, where it was found that the polar condensate is the stable ground state for $c_2>0$. 
In the opposite limit of extremely low particle density, and when $g_0=c_0-2c_2<0$, we demonstrate, that the ground state is a condensate of weakly interacting molecules, each being a bound state of two atoms in a spin-singlet state. Such a condensate does not break the $SU(2)$ symmetry, and its spin-excitations have a spectral gap equal to the binding energy of a molecule. 
 In  contrast to the metastable quasi-bound molecular states, appearing close to Feshbach resonance, as discussed in Ref. \cite{Timmermans99}, we find this phase thermodynamically stable against both  the collapse towards larger atomic clusters and the formation of the polar condensate due to  the breakup of molecular pairs. With increasing concentration, the molecules lose their individuality and the system becomes analytically intractable, in analogy with the BEC to BCS crossover in fermionic quantum gases \cite{BEC-BCS}. However, by continuity, the  $SU(2)$ symmetry persists until at some critical concentration, which roughly corresponds to the inter-atomic distance comparable with the size of a single molecule, the symmetry breaking phase transition to the polar condensate takes place. We would like to stress that we are only considering the zero temperature case.  At finite temperatures the situation is more complex due to the effects of long range thermal fluctuations \cite{Moore06,Lamacraft11}.

%

We begin our analysis with recalling some peculiarities of the Hamiltonian Eq. (\ref{Hamiltonian}) in two dimensions. The strength of the two-body interactions is determined by two coupling constants $m g_i/\hbar^2$ ($i=0,2$), which are dimensionless. This implies that at the classical level the Hamiltonian does not have an intrinsic length scale, and hence the symmetry of the mean field ground state is insensitive to the particle density. Beyond the mean field,  perturbative quantum corrections to the two-particle scattering amplitudes experience ultraviolet logarithmic divergences, amenable to the renormalization group treatment \cite{WilsonRG,Transmutation}. 
Thus, in the quantum picture, the parameters $g_0$ and $g_2$ should be considered as renormalized coupling constants, which depend on the running energy scale $E$ as \cite{Petrov2D}
\begin{equation}
g_i(E)=\frac{4\pi\hbar^2}{m}\frac{1}{\ln(E_i/E)}, 
\label{gE}
\end{equation}
where $E_i$ are determined by the ultraviolet (UV) physics. In Eq. (\ref{gE}) it is assumed that the running energy $E$ is less than the UV energy cutoff $\Lambda$. 
In the mean field theory, the energy scale $E$ is set by the chemical potential, which in the case of the polar condensate is given by $\mu=c_0 n$, where $n$ is the particle density \cite{Ho98}. Therefore, on the quantum level, the coupling constants in the Hamiltonian Eq. (\ref{Hamiltonian}) depend on the particle density. 

The mean field treatment of the Hamiltonian Eq. (\ref{Hamiltonian}) is justified for small coupling constants. 
In the case of weak repulsive interaction,  $\Lambda \ll E_i$, the coupling constants decrease with decreasing density. In contrast, any weak  attractive coupling ($\Lambda\gg E_i$, $g_i<0$) increases with decreasing energy, driving the Hamiltonian into a strong coupling regime.  Thus, while at high densities the Hamiltonian can still be analyzed within the mean field theory approach, in the low density limit the mean field theory breaks down. As is often the case, in such a limit the fundamental degrees of freedom can no longer be used for a meaningful description of the system's properties. Rather one has to work with the weakly interacting physical degrees of freedom, if those can be identified. 
In order to understand what those are in the present case, we first consider dilution so extreme, that only two atoms are present in the system. Since any attractive two particle interaction in two dimensions creates a bound state, the attraction in the singlet channel, $g_0<0$, binds the two atoms into a spin-singlet molecule. Such a molecule is characterized by its binding energy $E_0$, and its size $d=\sqrt{m E_0/\hbar}$. At small but finite  particle density, $n\ll\hbar/(mE_0)$, such molecules form a weakly nonideal Bose gas, in which  intermolecular collisions occur only rarely. The effect of such collisions depends on the sign of the intermolecular  interaction. For repulsive interaction, a stable  $SU(2)$ symmetric Bose-Einstein condensate of molecules is formed. Such a condensate is described by a scalar Gross-Pitaevskii functional with a running interaction constant $c_M (E)>0$, and with the UV cutoff determined by the binding energy of a molecule.  The energy of the molecular condensate is given by  $E_M=N (-E_0+c_M n/4)/2$, where $N$ and $n$ denote the total number and the density of atoms respectively. The molecular pair condensate is thermodynamically stable if $E_M$ is less than the energy of the polar condensate given by $E_p=c_0 n/2$. 
Obviously, this stability criterion is fulfilled for low atomic density.  
In contrast, intermolecular attraction, $c_M<0$, leads to the instability of the system against the collapse. 
In the rest of the paper, we focus on the detailed quantitative analysis  of intermolecular interaction, establishing the conditions under which the latter is  repulsive.



For the quantitative description of the intermolecular scattering we employ the Skorniakov and Ter-Martirosian (STM) formalism, see Refs. \cite{STM,Petrov}. This formalism is effective for determining the scattering amplitudes and the energies of the bound states in few-body scattering problems with  short range scattering potentials. The STM approach is based on the observation that in a system with a short range two-body interaction potential, many-particle dynamics is fully determined by the two-particle scattering amplitudes in the s-wave channel. 
The latter in turn can be emulated with the  Bethe-Peierls (BP) boundary condition imposed on the many-body wave function 
\begin{equation}
\frac{1}{\psi}\frac{\partial\psi}{\partial r_{ij}}\bigg\vert_{r_{ij}=R_0}=-\frac{1}{a}. 
\label{BP2D}
\end{equation}
Here $r_{ij}$ is the distance between the colliding particles $i$ and $j$. The parameter $R_0$ is related to the UV energy cutoff $\Lambda=\hbar^2/(m R_0^2)$.
 The scattering amplitude at energy $E$ only  depends on the dimensionless ratios $a/R_0$ and $E/\Lambda$, which reflects the renormalization group symmetry of the system in two dimensions.  The set of points satisfying $r_{ij}=R_0$ defines a hyper-cylinder in the  $2N$ dimensional configuration space. The union of such hyper-cylinders  for all pairs $i,j$ is called the  scattering surface, which we denote by $\mathcal{S}$.   
The formal solution of the $N$-particle scattering problem can be written in the form \cite{Petrov}
\begin{equation}
\psi_E({\bf r}) =\psi^0_{E}({\bf r})+\int_{\mathcal{S}} d{\bf r'} G_E({\bf r}-{\bf r'}) f_E({\bf r'}).     
\label{genSolution}
\end{equation}
Here $\psi_E({\bf r})$ denotes the exact many-particle wave functions at a given energy, $\psi^0_{E}({\bf r})$ denotes the incoming  wave, $G_E({\bf r}-{\bf r'})$ denotes the Green function, describing propagation of particles without collisions, and $f_E({\bf r'})$ is an auxiliary function defined on the scattering surface. 
In the STM formalism one applies the BP boundary condition Eq. (\ref{BP2D}) to Eq. (\ref{genSolution}), which results in a closed integral equation for the function $f_E({\bf r'})$. The latter is also called the STM equation. 

In the case of two particles, the function $f_E({\bf r})$ has no spacial dependence, and the explicit solution of the STM equation  for $E>0$ reads \cite{Supplement}   
\begin{equation}
f_E=\frac{1}{2\mu R_0\left[\ln\left(\sqrt{2\mu E} R_0 e^{a/R_0}\right)-i\pi/2\right]},   
\label{f_R>}
\end{equation}
where $\mu$ denotes the reduced mass. Analytical continuation of the solution Eq. (\ref{f_R>}) to the upper half-plane of complex energy results in a pole at the negative energy axis at  
\begin{equation}
E=-E_0=-\frac{1}{2\mu R_0^2}e^{-2a/R_0}.  
\label{BindingEnergy}
\end{equation}
This pole correspond to the formal solution of Eq. (\ref{genSolution}) without the incoming wave, which corresponds to a bound state at energy $-E_0$. 
For the scattering parameter $a>0$, the energy $E_0$ is much less than the high-energy cutoff $\Lambda$,  therefore  the pole corresponds to a bound state at energy $-E_0$.

It is instructive to relate the effective parameters $a$ and $R_0$ to the microscopic parameters of the trapped Bose condensate. The two-body scattering problem for a parabolically confined two-dimensional system was explicitely solved in Refs.  \cite{Petrov2D,Idziaszek}. In this system, the UV length scale is set by the oscillator length of the parabolic confinement potential $\ell_0$.   Comparing the results of \cite{Petrov2D,Idziaszek} with the scattering amplitude obtained from the Bethe-Peierls boundary condition Eq. (\ref{BP2D}), and setting for convenience the UV length cutoff $R_0=1.31 \ell_0$, we find  
\begin{equation}
\frac{a}{R_0}=-\sqrt{\frac{\pi}{2}}\frac{\ell_0}{a_{\mathrm{3D}}}, 
\label{aBP_a3D}
\end{equation}
where $a_{\mathrm{3D}}$ is the 3-dimensional s-wave scattering length. 
For the attractive interaction in the singlet channel, $a_{\mathrm{3D}}^{(0)}<0$ in Eq. (\ref{aBP_a3D}), we deduce that there is a bound state of two atoms, which is  a singlet pair with the binding energy $E_0$, as given by Eq. (\ref{BindingEnergy}).  
Comparing the intermolecular distance with the size of a molecule, we estimate the critical density    
\begin{equation}
n_c\sim \frac{1}{\ell_0^2} e^{\pi \ell_0/a_{3\mathrm{D}}^{(0)}} 
\end{equation}
of the phase transition between the polar and the singlet condensates.



Next we proceed to the calculation of the effective interaction between two singlet molecules. 
To this end, we  solve the two-pair scattering problem.   Since we consider only the scattering events without the dissociation of the pairs, the final state still consists of the two singlet molecules.  Moreover, the total spin of the intermediate four-particle state equals 0.  Guided by that reason, we introduce the basis of two-pair states in the spin space as follows 
$\Phi_i= |i,4\rangle_s\otimes|j,k\rangle_s$, where $i\neq j\neq k\neq 4$. Each state consists of two singlet pairs of atoms, and it is marked by the number of the atom that forms a singlet with the atom 4. 
The general two-pair wave function reads 
\begin{equation}
\Psi({\bf r}_1, {\bf r}_2, {\bf r}_3, {\bf r}_4)= \sum_{i=1}^3\chi_i({\bf r}_1, {\bf r}_2, {\bf r}_3, {\bf r}_4) \Phi_i. 
\label{4Pwf}
\end{equation}
Here $\chi_i$ describes the spatial part of the wave function and $\Phi_i$ relates to the spin part. Furthermore, we separate the center of mass coordinate of the four atoms, and introduce the following set of Jacobi-coordinates to describe their relative position 
\begin{equation}
{\bf z}=({\bf r}_3-{\bf r}_4), \, {\bf y}=({\bf r}_1-{\bf r}_2), \,  {\bf x}=\frac{1}{\sqrt{2}}[({\bf r}_3+{\bf r}_4)-({\bf r}_1+{\bf r}_2)].  \label{relative_coordinates}
\end{equation}
Since we have two scattering channels (spin 0 and spin 2), there are two sets of BP boundary conditions that read 
\begin{equation}
 \partial_{r_{ij}}\hat{P}_{ij}^{(\nu)}\Psi\big\vert_{r_{ij}=R_0}=-\frac{1}{a_{\nu}} \hat{P}_{ij}^{(\nu)}\Psi\vert_{r_{ij}=R_0}.
\label{BP2D_twochannels}
\end{equation}
Here $r_{ij}=|{\bf r}_i-{\bf r}_j|$, $\hat{P}_{ij}^{(\nu)}$  denotes the projection operator on the subspace, in which the atoms $i, j$ have the total  spin $\nu$, $(\nu=0,2)$. Applying Eq. (\ref{BP2D_twochannels}) to Eq. (\ref{genSolution}), we obtain the set of equations for  functions $f_{ij}^{\nu}$, each having its domain on the  hyper-cylinder $r_{ij}=R_0$.  The symmetry of the bosonic wave function under the permutations of atoms further reduces the number of independent unknown functions to two,  $f_0({\bf x}, {\bf z})$ and $f_2({\bf x}, {\bf z})$,  which are related to the scattering amplitudes in the $S=0$ and $S=2$ channels respectively. The variable ${\bf y}$ has been eliminated by BP boundary conditions applied on the hyper-cylinder $|{\bf r}_1-{\bf r}_2|=|{\bf y}|=R_0$.  After the Fourier transform in the variables ${\bf x}$ and ${\bf z}$, the STM equations for the two-channel problem take the form \cite{Supplement}
\begin{eqnarray}
\nonumber && 
\alpha_{\nu}({\bf k, p}) f_{\nu}({\bf k, p})=\frac{2\sqrt{\pi}(2-\epsilon)\delta_{\nu, 0}}{1+p^2(2-\epsilon)}\delta({\bf k}-{\bf K}) +\\ 
&& 
\int\frac{d^2{\bf Q}}{(2\pi)^2} \frac{2}{1+k^2+p^2+Q^2} 
\left\{ f_{\nu}(-{\bf k}, {\bf Q})+ 
 2\mathcal{K}_{\nu\mu}\sum_{s=\pm}  f_{\mu}\left(\frac{{\bf p}+{\bf Q}}{\sqrt{2}}, s\frac{{\bf Q}-({\bf k}+{\bf p})}{\sqrt{2}}\right) 
\right\}.
\label{Eqf_general}
\end{eqnarray} 
Here  $ \epsilon = (E +2E_0)/E_0$ is the dimensionless energy relative to the rest energy of two isolated molecules.   The energy $E_0$ is given by Eq. (\ref{BindingEnergy}) for the singlet scattering channel,   $a=a_0$. The Fourier conjugates of  ${\bf x}$ and  ${\bf z}$, which we denote as ${\bf k}$ and ${\bf p}$ respectively,  are measured in units of $\sqrt{|E|}$. The functions $\alpha_{\nu}({\bf k, p})$ are defined by 
\begin{equation}
\alpha_{\nu}({\bf k, p}) =\frac{1}{2\pi} \ln[(2-\epsilon)(1 + k^2+p^2)]-\lambda\delta_{\nu, 2}, 
\end{equation}
the subscripts $\mu, \nu = 0,2$ relate to the scattering channels. The scattering in the spin-2 scattering channel enters Eq. (\ref{Eqf_general}) through the parameter $\lambda$, which is defined as 
\begin{equation}
\lambda=\frac{(a_0+|a_2|)}{\pi R_0}=\frac{\ell_0}{\sqrt{2\pi}}\left(\frac{1}{|a_{\mathrm{3D}}^{(0)}|}+\frac{1}{|a_{\mathrm{3D}}^{(2)}|}\right).
\label{lambda}
\end{equation}
where the second equation is obtained in virtue of Eq. (\ref{aBP_a3D}), and we consider the case $a_{\mathrm{3D}}^{(0)}<0$, corresponding to the attraction in the singlet channel.  The first term on the RHS of Eq. (\ref{Eqf_general}) represents  the incoming wave with the relative momentum of the two molecules ${\bf K}$. We note, that the incoming wave is only present in the singlet channel, $\nu=0$.  
The explicit form of the matrix $\mathcal{K}_{\nu\mu}$ follows from the symmetry of the wave function by spin rotations and permutations of atoms, it reads 
\begin{equation}
\mathcal{K}=\left(\begin{array}{cc}
1/3 & 5/9 \\
1 & 1/6
\end{array}\right).
\label{MatrixK}
\end{equation}

If the interaction between two molecules is attractive, then, by analogy with Eqs. (\ref{f_R>}), (\ref{BindingEnergy}), the analytical continuation of $f_{\nu}(\epsilon)$  acquires a pole for $\epsilon<0$, $|\epsilon|\ll 1$, which corresponds to a formation of two-molecule bound complex. At the energy of the bound state there is a nontrivial solution of STM equations without  the incoming wave.  It may also happen that the combination of attraction in the $S=0$ channel and repulsion in the $S=2$ channel results in the overall repulsion between two singlet pairs. In that case no four-particle bound state should exist. 
\begin{figure}
\includegraphics[width=8cm]{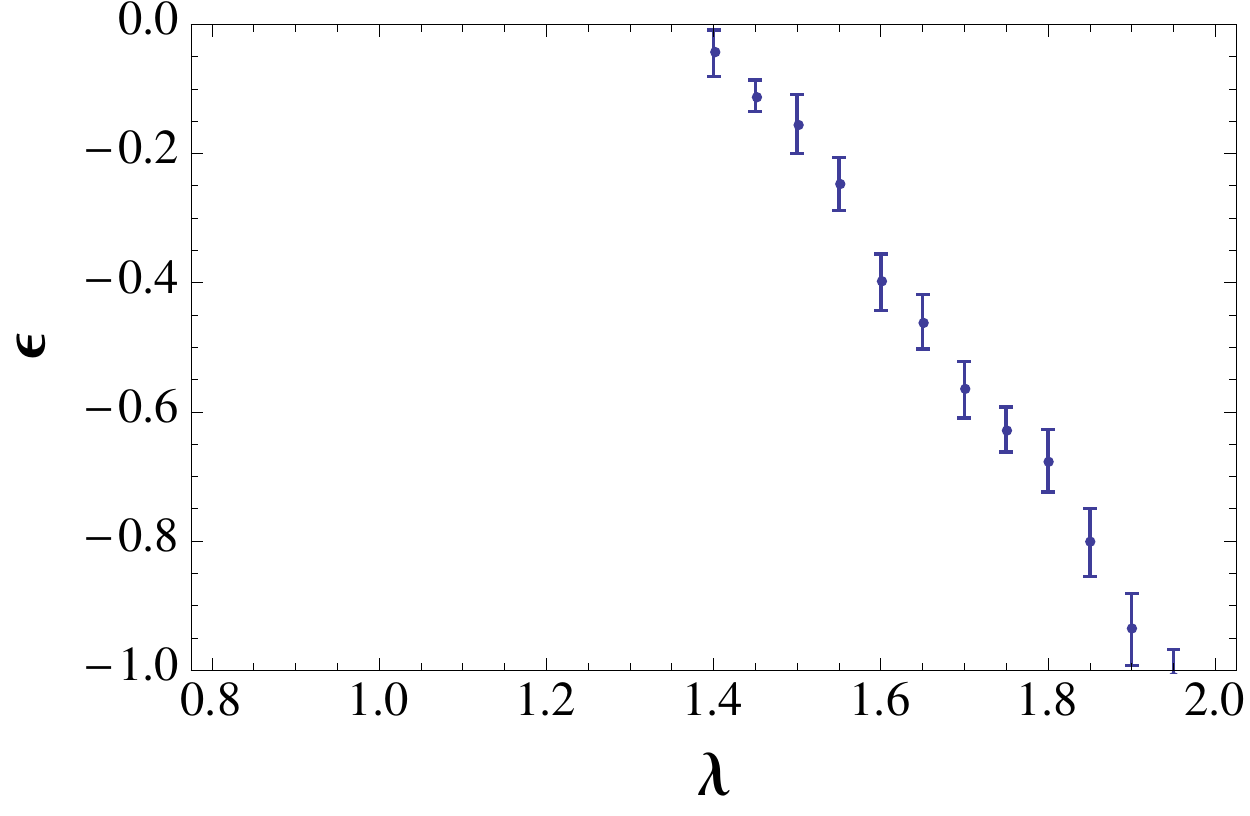}%
\caption{Energy $\epsilon$ of the stationary state of Markovian evolution, corresponding to the bound state of two molecules with energy $-2+\epsilon$ (in units of $E_0$) as a function of parameter $\lambda$ measuring the relative strength of attraction in spin-0 and repulsion in spin-2 channels (see Eq. (\ref{lambda})), averaged over 20 independent runs of Markovian evolution. 
The two-molecule bound state appears for $\lambda\approx 1.4\pm 0.1$ as a solution with negative $\epsilon$, which determines the boundary for the stability of the SMC ground state.  
\label{fig-Lambda_Epsilon}}
\end{figure}
We solve Eq. (\ref{Eqf_general}) numerically, using the method of stochastic Markovian evolution with branching \cite{Monte-Carlo}. 
The energy of the bound state is found by adjusting the value of $\epsilon$, such that the overall number of walkers fluctuates around a time-independent mean value \cite{Supplement}.  
Numerical results are shown in Fig. \ref{fig-Lambda_Epsilon}.  
Decreasing $\lambda$, which corresponds to increase of repulsive interactions in the spin-2 channel (see Eq. (\ref{lambda})) reduces the  absolute value of $\epsilon$, at which a stationary solution is found, and below $\lambda_c=(1.3\pm 0.1)$ no stationary solution exists, which means that the 4-atomic bound state disappeared from the spectrum of the system. Therefore, for $\lambda <\lambda_c$ the condensate of singlet pairs is stabilized against further collapse.  Eq. (\ref{lambda}) allows to reformulate the condition $\lambda < 1.3$ in terms of the three-dimensional scattering lengths and the oscillator length of confinement potential
\begin{equation}
\frac{\ell_0}{|a_{\mathrm{3D}}^{(2)}|}<1.3 \sqrt{2\pi}-\frac{\ell_0}{|a_{\mathrm{3D}}^{(0)}|}. 
\label{MainCondition}
\end{equation}
Eq. (\ref{MainCondition}) specifies the conditions for the three-dimensional scattering length in $S=0$ and $S=2$ channels, and the characteristic length of the confinement potential, at which the two-dimensional condensate of singlet pairs is stable against collapse. Experimentally, the regime specified by Eq. (\ref{MainCondition}) can be realized with help of optical Feshbach resonance \cite{Fedichev,Bohn97,Bohn99} or using the radio-frequency dressed atomic states \cite{Cheianov-Chudnovskiy}. Creation of the SMC state opens the way to  further investigation of the wide variety of exotic phases that have been outlined in the beginning of this paper.

\begin{acknowledgments}
Authors acknowledge the hospitality of the Institute for Basic Science in Daejeon, Korea, where part of this work has been performed. 
\end{acknowledgments}

\noindent{\bf Data Availability}\\
The Mathematica notebook file generating all data analyzed during this study is available under the link \\
https://drive.google.com/open?id=11wdTyWEzRgTX8B2DEqJe9jURr7z9o0JV.

\newpage
\section{Supplementary material}

In this supplementary material we provide details of the derivation of the main formulas in the text of the paper. 

\subsection{Markovian evolution}
In this section we describe the mapping of Eq. (13) in the main text on the Markovian evolution process. For the sake of consistency, we repeat here Eq. (13) without incoming wave term 
\begin{eqnarray}
\nonumber &&
\left[\frac{1}{2\pi} \ln[(2-\epsilon)(1 + k^2+p^2)]-\lambda\delta_{\nu 2}\right]f_{\nu}({\bf k}, {\bf p})=\\ 
\nonumber && 
\int\frac{d^2{\bf Q}}{(2\pi)^2} \frac{2}{1+k^2+p^2+Q^2} 
\left\{ f_{\nu}(-{\bf k}, {\bf Q})+ 
 2\sum_{\mu=0,2} \mathcal{K}_{\nu\mu}\sum_{j=0,1}  f_{\mu}\left(\frac{{\bf p}+(-1)^j{\bf Q}}{\sqrt{2}}, \frac{{\bf Q}-(-1)^j({\bf k}+{\bf p})}{\sqrt{2}}\right) 
\right\} \\
\label{start}
\end{eqnarray}
with 
\begin{equation}
\mathcal{K}=\left(\begin{array}{cc}
1/3 & 5/9 \\
1 & 1/6
\end{array}\right).
\label{MatrixKsup}
\end{equation}
Let us  introduce functions $g_{\nu}({\bf k})$ defined as 
\begin{equation}
g_{\nu}({\bf k}, {\bf p})=f_{\nu}({\bf k}, {\bf p})\frac{\ln[(2-\epsilon)(1+k^2+p^2)]}{1+k^2+p^2}.
\label{gauge-transform}
\end{equation} 
After the transformation Eq. (\ref{gauge-transform}), equations for $g_{\nu}({\bf k})$ are written in the form that allows their iterative solution 
\begin{equation}
g_{\nu,n+1}({\bf k}, {\bf p})=\sum_{\mu=0,2} \int d^2{\bf k}' d^2{\bf p}'  P_{\nu\mu}({\bf k}, {\bf p}; {\bf k}', {\bf p}') g_{\nu,n}({\bf k}', {\bf p}')
\label{g_iterative}
\end{equation} 
It is important to note, that the gauge transformation Eq. (\ref{gauge-transform}) ensures that the integrals of the kernels $P_{\nu\mu}({\bf k}, {\bf p}; {\bf k}', {\bf p}')$ over the first coordinates ${\bf k}, {\bf p}$ are finite, which allows the interpretation of $P_{\nu\mu}({\bf k}, {\bf p}; {\bf k}', {\bf p}')$ as a transition rate from the state $|{\bf k}',{\bf p}', \mu\rangle$ to the state  $|{\bf k}, {\bf p}, \nu\rangle$, and rewriting Eq. (\ref{g_iterative}) in form of a master equation 
\begin{eqnarray}
\nonumber && 
g_{\nu,n+1}({\bf k}, {\bf p})-g_{\nu,n}({\bf k}, {\bf p})=\\ 
&& 
\sum_{\mu=0,2} \int d^2{\bf k}' d^2{\bf p}' \left[P_{\nu\mu}({\bf k}, {\bf p}; {\bf k}', {\bf p}')-\delta_{\nu\mu}\delta({\bf k}-{\bf k}')\delta({\bf p}-{\bf p}')\right]  g_{\nu,n}({\bf k}', {\bf p}').
\label{g_master}
\end{eqnarray}
The kernels $P_{\nu\mu}({\bf k}, {\bf p};  {\bf k}', {\bf p}')$ are obtained straightforwardly from Eq. (\ref{start}) and transformation Eq. (\ref{gauge-transform}).   
The bound four-atomic state is realized as a stationary solution of Eq. (\ref{g_master}). 
To realize the numerical implementation of Eq. (\ref{g_iterative}) as a stochastic Markovian evolution process, we need to interpret the transition rates in Eq. (\ref{g_master}) as {\em probabilities} of a jump of a particle. To reach this, we divide RHS of Eq. (\ref{g_master}) by a maximal value of the total escape rates 
\begin{equation}
\Gamma_{\mu}({\bf k}', {\bf p}')=\sum_{\nu=0,2}\int d^2{\bf k} d^2{\bf p} P_{\nu\mu}({\bf k}, {\bf p}; {\bf k}', {\bf p}')
\end{equation} 
from the state $|{\bf k}', {\bf p}', \mu\rangle$. This operation is equivalent to the rescaling of the (discrete) time in Eq. (\ref{g_master}), thus it does not change the stationary state we are interested in. The resulting equations read 
\begin{eqnarray}
\nonumber && 
g_{\nu, n+1}({\bf k}, {\bf p})-g_{\nu, n}({\bf k}, {\bf p})=-\gamma_{\nu}({\bf k}, {\bf p}) g_{\nu,n}({\bf k}, {\bf p})+\\ 
&& 
\sum_{\nu'}\int d^2{\bf k}'  d^2{\bf p}'\left\{ W_{\nu\nu'}({\bf k}, {\bf p};  {\bf k}', {\bf p}') g_{\nu', n}({\bf k}', {\bf p}')
-W_{\nu'\nu}({\bf k}', {\bf p}'; {\bf k}, {\bf p})g_{\nu, n}({\bf k}, {\bf p})\right\}, 
\label{matrix_Master-Eq}
\end{eqnarray}
where 
\begin{equation}
W_{\nu\nu'}({\bf k}, {\bf p}; {\bf k}', {\bf p}')=P_{\nu\nu'}({\bf k}, {\bf p}; {\bf k}', {\bf p}')/C,
\end{equation} 
\begin{equation}
\gamma_{\nu}({\bf k}, {\bf p})=\frac{1}{C}\left(1-\Gamma_{\nu}({\bf k}, {\bf p})-\frac{\lambda}{\ln[(2-\epsilon)(1+k^2+p^2)]}\delta_{\nu 2}\right), 
\label{gamma_k}
\end{equation}
and 
\begin{equation}
C>\max_{\{{\bf k}, {\bf p}\}}\left\{\Gamma_{2}({\bf k}, {\bf p})+\frac{\lambda}{\ln[(2-\epsilon)(1+k^2+p^2)]}, \Gamma_{0}({\bf k}, {\bf p})\right\}.
\label{C}
\end{equation}
For practical calculations in the region $-1<\epsilon\leq 0$, $0<\lambda<2$, $C=20$ is the optimal choice. 
The choice of the factor $C$ garanties 
\begin{equation}
\sum_{\nu=0,2}\int d^2{\bf k}d^2{\bf p} W_{\nu\nu'}({\bf k}, {\bf p}; {\bf k}', {\bf p}')  <1,
\end{equation} 
which allows the interpretation of $W_{\nu\nu'}({\bf k}, {\bf p}; {\bf k}', {\bf p}')$ as a probability density for the jump of the particle out of the state $|{\bf k}', {\bf p}', \nu'\rangle$ into the state $|{\bf k}, {\bf p}, \nu\rangle$. 
Now we can formulate Markovian stochastic process, which is described by the master equation Eq. (\ref{matrix_Master-Eq}) as follows: Consider a ensemble of walkers that evolve in the 4-dimensional space ${\bf k}=({\bf k}, {\bf p})$ and have an intrinsic flavor $\nu=0,2$. At each discrete time step a walker in the state $|{\bf k}, {\bf p},\nu\rangle$ is subject to the following elementary process: (i) jump to the state $|{\bf k'},{\bf p}',\nu'\rangle$ with the probability $W_{\nu'\nu}({\bf k'}, {\bf p}'; {\bf k}, {\bf p})$ by changing the flavor to $\nu'$ (the same flavor is kept if $\nu'=\nu$); (ii) the walker is destroyed with the probability $\gamma_{\nu}({\bf k}, {\bf p})$. If $\gamma_{\nu}({\bf k}, {\bf p})<0$, another walker is created in the state $|{\bf k}, {\bf p},\nu\rangle$ with the probability $|\gamma_{\nu}({\bf k}, {\bf p})|$. 

Implementing that algorithm numerically, the stationary solution is distinguished by a  total number of walkers fluctuating around a stable mean value. Generically, due to finite probabilities $\gamma_{\nu}({\bf k})$ either all initially created walkers die out, or their number grows unbound. The final outcome of the evolution is crucially affected by the term $\lambda/\ln(2-\epsilon+k^2+p^2)$ that governs creation or annihilation of walkers in the $\nu=2$ channel. For a generic value of $\lambda$, the total  number of walkers grows without bounds for small $|\epsilon|$, and decays to zero after $|\epsilon|$ exceeds some critical value, that corresponds to the energy of the bound 4-atomic state. In the numerical procedure, the value of $\epsilon$ is adjusted to reach the situation with stationary average number of walkers.  

\subsection{Two particle scattering amplitude from the Bethe-Peierls boundary conditions in two dimensions}

Consider a collision of two particles with mass 1 in 2D. We are interested in s-wave scattering.  
The wave function as a function of the  the relative coordinate satisfies the Schr\"odinger equation 
\begin{equation}
 \left\{\left(\frac{d^2}{dr^2}+\frac{1}{r}\frac{d}{dr}\right) + 2\mu(E-V(r))\right\}\psi(r)=0.
\label{Schroedinger_1}
\end{equation}
Here $\mu$ is the reduced mass, $\mu=1/2$.  If $V(r)$ is a deep finite range potential, it can be emulated by the Bethe-Peierls boundary condition, which in 2D is formulated on a circle of small compared to the de Broglie wave length radius $R_0$. The Bethe-Peierls boundary condition reads 
\begin{equation}
\frac{d\psi/dr}{\psi}\bigg\vert_{r=R_0}=-\frac{1}{a}. 
\label{BP2Dsup}
\end{equation}

The solution of the scattering problem can be written in terms of the Green function describing a free motion of particles in 2D space as follows 
\begin{equation}
\psi({\bf r}, t) =\psi_0({\bf r}, t)+\int d^2{\bf r'} G({\bf r}-{\bf r'})\delta(|{\bf r}-{\bf r'}|-R_0) f({\bf r'}). 
\label{genSolutionsup}
\end{equation}
The function $\psi_0$ describes the incoming plane wave. The Green function describes the scattered wave, and the integration goes over the region, where the scattering takes place. The Green function satisfies the equation 
\begin{equation}
\left(-\frac{1}{2\mu}\nabla^2 - E\right) G({\bf r}, {\bf r'}) =\delta({\bf r}-{\bf r'}). 
\label{GF}
\end{equation}
Introduce the parameter $k=\sqrt{2\mu |E|}$, and the rescaled coordinate ${\bf x}=k{\bf r}$. Then for $E>0$ the scattered wave is described by the function $G(|{\bf r}-{\bf r'}|)=C H_0^{(1)} (|{\bf r}-{\bf r'}|)$, where $H_0^{(1)}(x)$ denotes the Hankel function. The constant $C$ is determined by substitution in Eq. (\ref{GF}) and integrating over the circle $|{\bf r}-{\bf r'}|=R_0$. Using the short range approximation $H_0^{(1)}(x)\approx\frac{2 i}{\pi} \ln x+1$, we get $C=\frac{i\mu}{2}$ and hence 
\begin{equation}
G(x)=\frac{i\mu}{2} H_0^{(1)}(x). 
\end{equation}
For $E<0$ the scattered wave is described by $G(x)= C K_0(x)$, where $K_0(x)$ is the modified Bessel function of the second kind. Using the short-range asymptote $K_0(x)\approx -\ln x$, we fix the constant $C=\mu/\pi$ and obtain ($x=\sqrt{2\mu|E|} r)$,  $E<0$ 
\begin{equation}
 G_E(x)=\frac{\mu}{\pi} K_0(x).      
 \label{GE_x}
\end{equation}
Now we fix the function $f({\bf r})$ by substituting the formal solution Eq. (\ref{genSolution}) in the boundary condition Eq. (\ref{BP2D}). Thereby the action of the scattering potential is replaced by a function $f({\bf r'})$ on the circle of a small radius $R_0$. We obtain two equations, the one for the wave function and the one for its derivative 
\begin{eqnarray}
\psi({\bf r}, t) =\psi_0({\bf r}, t)+\oint_{|r'|=R_0} d{\bf l'} G({\bf r}-R_0) f(R_0), \label{psi}\\
\partial_r\psi({\bf r}, t) =\partial_r\psi_0({\bf r}, t)+\oint_{|r'|=R_0} d{\bf l'} \partial_r G({\bf r}-R_0) f(R_0). \label{dr_psi}
\end{eqnarray}
s-wave scattering implies the angular independence of $f(R_0)$, which allows to put it out of the integration. Furthermore, at small distances (correspondingly large wave vectors), one can neglect the energy in Eq. (\ref{GF}) for the Green function. In that way the equation for the Green function acquires the form of the equation for the Coulomb potential in 2D. Putting $\mu=1/2$, we get 
\begin{equation}
\nabla^2 G({\bf r}, {\bf r'}) =-\delta({\bf r}-{\bf r'}).  
\end{equation}
The condition for that approximation reads 
\begin{equation}
\sqrt{2\mu |E|} R_0\ll 1, 
\end{equation}
which determines the small parameter in the following derivations. 
Furthermore, using the Gauss theorem, we can understand the integral over the circle as a potential created by the homogeneous charge distribution on the circle, which in turn is equal to the potential of the total charge places in the center of the  circle. It follows that the result of the integration does not change if  we replace the argument $R_0$ by zero in the Green function. Applying this line of arguments we obtain 
\begin{equation}
\oint_{|r'|=R_0} G_E({\bf r}-{\bf r'}) f(R_0) d{\bf l} = f(R_0)\oint_{|r'|=R_0} G_E({\bf r}-{\bf R_0}) d{\bf l}=
2\pi R_0 f(R_0) G_E(r). 
\label{Coulomb}
\end{equation}
Now we can evaluate integrals in in Eqs. (\ref{psi}), (\ref{dr_psi}) using Eq. (\ref{Coulomb}). Furthermore, since the derivative of the incoming wave $\partial_r\psi_0$ is a smooth function at small $r$ whereas the Green function develops a singularity, one can neglect the term $\partial_r\psi_0$ in Eq. (\ref{dr_psi}). Then the boundary condition assumes the form 
\begin{equation}
 \frac{\partial_r\psi}{\psi}\vert_{r=R_0}=\frac{2\pi R_0 f(R_0)\partial_r G_E(R_0)}{\psi_0(R_0)+2\pi R_0 f(R_0) G_E(R_0)} =-\frac{1}{a}. 
 \label{BC1}
\end{equation}
Using the asymptote of Green function at small $r$, and replacing $\psi_0(R_0)\approx 1$, we solve Eq. (\ref{BC1}) with respect to $f(R_0)$. The solution the form 
\begin{eqnarray}
&& 
 f(R_0)=\frac{1}{2\mu R_0\left[\ln\left(k R_0 e^{a/R_0}\right)-i\pi/2\right]},  \  \mbox{for}  \ E>0,  \label{f_R>sup}\\
&& 
 f(R_0)=\frac{1}{2\mu R_0\ln\left(k R_0 e^{a/R_0}\right)} , \  \mbox{for}  \  E<0.
 \label{f_R<}
 \end{eqnarray}
Furthermore, in the case $E>0$ we can relate $f(R_0)$ to the scattering amplitude. Substituting $f(R_0)$ in the general solution Eq. (\ref{genSolution}), we obtain for $E>0$ 
\begin{equation}
\psi({\bf r}, t) =\psi_0({\bf r}, t)+i\pi \mu R_0 f(R_0) H_0^{(1)}(kr).  
\label{scattered_wave1}
\end{equation}
Using the large distance asymptote 
\begin{equation}
 H_0^{(1)}(x)=\sqrt{\frac{2}{\pi x}}e^{ix}e^{-i\pi/4}
\end{equation}
we write down Eq. (\ref{scattered_wave1}) in the form 
\begin{equation}
 \psi({\bf r}, t) =\psi_0({\bf r}, t)+e^{i\frac{\pi}{4}} \sqrt{\frac{2\pi}{k}} \mu R_0 f(R_0) \frac{e^{ikr}}{\sqrt{r}},   
\label{scattered_wave}
\end{equation}
from which we identify the scattering amplitude as 
\begin{equation}
A=\sqrt{\frac{2\pi}{k}} \mu R_0 f(R_0)=\frac{\sqrt{\pi}}{\sqrt{2k}\left[\ln\left[k R_0  e^{\frac{a}{R_0}}\right]-i\pi/2\right]},  
\label{scatt_amplitude_1}
\end{equation}
where we put $\mu=1/2$. 
The continuation of the scattering amplitude to negative energies is obtained by using the expression (\ref{f_R<}) in Eq. (\ref{scatt_amplitude}), which results in 
\begin{equation}
A=\frac{\sqrt{\pi}}{\sqrt{2k}\ln\left[k R_0 e^{\frac{a}{R_0}}\right]},  
\label{scatt_amplitude}
\end{equation}
For $a>0$, the scattering amplitude has a  pole as a function of the energy ($k=\sqrt{|E|}$) at 
\begin{equation}
E=-|E|=-\frac{1}{R_0^2}e^{-2\frac{a}{R_0}}, 
\end{equation}
which corresponds to the formation of a bound molecular state. 
For $a<0$, the scattering amplitude remains negative for all energies without showing any resonant structure. 

To obtain the relation of the scattering parameter $a$ with the s-wave scattering length in 3 dimensions, we compare Eq. (\ref{scatt_amplitude}) with the expressions for the scattering amplitudes in presence of confinement potential derived in Refs. \cite{Petrov2D,Idziaszek}. The comparison results in the following equation 
\begin{equation}
\ln\left[k R_0 e^{\frac{a}{R_0}}\right]=-\sqrt{\frac{\pi}{2}}\frac{\ell_0}{a_{\mathrm{3D}}} +\ln\left(k\ell_0\sqrt{\frac{\pi}{2B}}\right). 
\label{Comparison}
\end{equation}
Equating $k$-dependent and $k$-independent parts of Eq. (\ref{Comparison}), we obtain 
\begin{equation}
\frac{a}{R_0}=-\sqrt{\frac{\pi}{2}}\frac{\ell_0}{a_{\mathrm{3D}}}, \,  R_0=\ell_0\sqrt{\frac{\pi}{2B}}, 
\label{aBP_a3Dsup}
\end{equation}
which constitutes  Eq. (3) of the main text of the paper.

\subsection{Structure of the wave function for scattering of two singlet pairs}
We consider scattering of two singlet bound states, which we also call molecules.  The total spin of the four-particle state equals 0. We only consider the elastic scattering events, therefore the out-state still consists of the two singlet molecules. Guided by that reason, we introduce the basis $\mathcal{B}=\{\Phi_1, \Phi_2, \Phi_3\}$ in the spin-0 subspace of the spin Hilbert space of the four atoms as follows 
\begin{equation}
\Phi_1= |1,4\rangle_s\otimes|2,3\rangle_s, \, \, 
\Phi_2= |2,4\rangle_s\otimes|1,3\rangle_s, \, \, 
\Phi_3= |3,4\rangle_s\otimes|1,2\rangle_s.
\label{BasisB}
\end{equation}
Here $|i,j\rangle_s$ denotes the singlet state formed by the atoms $(i,j)$, the index of the state $\Phi_i$ corresponds to the  number of the atom that forms a singlet state with the atom 4. 
The general two-molecule wave function can now be written as 
\begin{equation}
\Psi({\bf r}_1, {\bf r}_2, {\bf r}_3, {\bf r}_4)= \chi_1({\bf r}_1, {\bf r}_2, {\bf r}_3, {\bf r}_4) \Phi_1 + \chi_2({\bf r}_1, {\bf r}_2, {\bf r}_3, {\bf r}_4) \Phi_2+ \chi_3({\bf r}_1, {\bf r}_2, {\bf r}_3, {\bf r}_4) \Phi_3.
\label{4Pwfsup}
\end{equation}
Here $\chi_i$ describes the spatial part of the wave function and $\Phi_i$ relates to the spin part, and ${\bf r}_j$, $j=1,2,3,4$ is the coordinate of the $j$'s atom. 

\subsection{Representation of permutation operators in the basis $\mathcal{B}$}
Direct calculation shows that in the basis  $(\Phi_1, \Phi_2, \Phi_3)$ the permutation operators are represented by 
\begin{equation}
 \Pi_{12}=\Pi_{34}=\left(\begin{array}{ccc}
              0 & 1 & 0 \\
	      1 & 0 & 0 \\
	      0 & 0 & 1 
                \end{array}\right), 
\end{equation}
\begin{equation}
 \Pi_{13}=\Pi_{24}=\left(\begin{array}{ccc}
              0 & 0 & 1 \\
	      0 & 1 & 0 \\
	      1 & 0 & 0 
                \end{array}\right), 
\end{equation}
\begin{equation}
 \Pi_{14}=\Pi_{23}=\left(\begin{array}{ccc}
              1 & 0 & 0 \\
	      0 & 0 & 1 \\
	      0 & 1 & 0 
                \end{array}\right). 
\end{equation}
Representations of other permutation operators are obtained according to the obvious relation $\Pi_{ij}=\Pi_{ji}$. 

\subsection{Projectors on the $F=0$ and $F=2$ scattering channels in the basis $\mathcal{B}$}
We denote $\hat{P}_{ij}^{(\nu)}$ the projection operator on the subspace, in which the atoms $i, j$ have the total  spin $\nu$, $(\nu=0,2)$. \\
 Explicit form of the projector onto $F=0$ state in terms of spin operators can be written as 
\begin{equation}
 \hat{P}_{ij}^{(0)}=\frac{1}{12}[(\hat{\bf S}_i+\hat{\bf S}_j)^2-6][(\hat{\bf S}_i+\hat{\bf S}_j)^2-2]. 
\label{Pij_0}
\end{equation}
Using Eq. (\ref{Pij_0}), and definition Eq. (\ref{BasisB}), we obtain the following matrix presentation of the projectors in the basis $\mathcal{B}$ 
\begin{equation}
 \hat{P}_{12}^{(0)}=\left(\begin{array}{ccc}
 0 & 0 & 0 \\
 0 & 0 &  0 \\
  \frac{1}{3} & \frac{1}{3}  & 1 
                          \end{array}
\right), \,  
\, \, 
\hat{P}_{23}^{(0)}=\left(\begin{array}{ccc}
1 & \frac{1}{3}  & \frac{1}{3}  \\
  0 & 0 & 0 \\
0 & 0 & 0  
                          \end{array}
\right), \,  
\, \, 
 \hat{P}_{31}^{(0)}=\left(\begin{array}{ccc}
0 & 0 & 0 \\
\frac{1}{3} & 1 &  \frac{1}{3}\\
0 & 0 & 0  
                          \end{array}
\right). 
\label{P_0}
\end{equation}
According to the construction of the basis states, the singlet state of the atom 4 and the atom $i$ means also the singlet of the two complementary atoms, $j$ and $k$, where $j,k \neq i, 4$. Therefore, for the projectors involving the atom 4, we have
\begin{equation}
\hat{P}_{14}^{(0)}= \hat{P}_{23}^{(0)},  \  \hat{P}_{24}^{(0)}= \hat{P}_{13}^{(0)}, \ \hat{P}_{34}^{(0)}= \hat{P}_{12}^{(0)}.
\end{equation}
Explicit form of the projector onto $F=2$ state in terms of spin operators can be written as 
\begin{equation}
 \hat{P}_{ij}^{(2)}=\frac{1}{24}(\hat{\bf S}_i+\hat{\bf S}_j)^2[(\hat{\bf S}_i+\hat{\bf S}_j)^2-2]. 
\label{Pij_2}
\end{equation} 
The matrix representation of the projection operator $\hat{P}_{12}^{(2)}$ is given by  
\begin{equation}
\hat{P}_{12}^{(2)}=\frac{1}{6}\left(\begin{array}{ccc}
                         3 & 3 & 0 \\ 
			  3 &3 & 0  \\ 
			  - 2 &  - 2 & 0
                         \end{array}\right). 
\label{P2-matrix}
\end{equation}
Other projection operators $P^{(2)}_{ij}$ are obtained by action of the permutation operator on $\hat{P}_{12}^{(2)}$ according to the rule 
\begin{equation}
P^{(2)}_{ij}=\Pi_{1i}\Pi_{2j}P^{(2)}_{12}\Pi_{1i}\Pi_{2j}
\end{equation}
 
\subsection{Derivation of STM equations}

Now the Bethe-Peierls boundary conditions can be formulated for each pair of particles $i,j$ as 
\begin{equation}
 \left(\partial_{r_{ij}}\hat{P}_{ij}^{(\nu)}\boldsymbol{\chi}+\frac{1}{a_{\nu}} \hat{P}_{ij}^{(\nu)}\boldsymbol{\chi}\right)\bigg|_{|{\bf r}_{ij}|=R_0}=0,   
\label{BP_chi}
\end{equation}
where $\boldsymbol{\chi}=(\chi_1, \chi_2, \chi_3)^T$.
The boundary conditions Eq. (\ref{BP_chi}) are formulated on the circle $r_{ij}=R_0$. 
The general solution in terms of Green's function can be written as (cf. Eq. (\ref{genSolution}))
\begin{equation}
\boldsymbol{\chi}=\boldsymbol{\chi}_0+\sum_{\langle i, j\rangle} \int_{|{\bf r}'_i-{\bf r}'_j|=R_0}G_E({\bf X}-{\bf X}') \boldsymbol{f}^{ij}({\bf X'}) d{\bf X'}, 
\label{genSol_3body}
\end{equation} 
where 
\begin{equation}
\left(\frac{1}{2}\nabla^2_{\bf X}+E \right)G_{E}({\bf X-X'})=- \delta({\bf X-X'}), 
\label{GF4-eq}
\end{equation}
and we introduced  8-dimensional coordinate vectors 
\begin{equation}
{\bf X}=({\bf r}_1, {\bf r}_2, {\bf r}_3, {\bf r}_4),  \, \, \, {\bf X}'=({\bf r}'_1, {\bf r}'_2, {\bf r}'_3, {\bf r}'_4). 
\label{X}
\end{equation}
In Eq. (\ref{genSol_3body})  we introduced a vector-valued function $\boldsymbol{f}^{ij}=(f^{ij}_1, f^{ij}_2, f^{ij}_3)^{T}$ for each pair of points $(ij)$ with its domain given by the cylinder $r'_{ij}=R_0$.  

The symmetry of the wave function under permutations of atoms induces linear relationships between the functions 
$\boldsymbol{f}^{ij}$. 
\begin{eqnarray}
&& 
\hat{\Pi}_{ij} \boldsymbol{f}^{ij}=\boldsymbol{f}^{ij}, \label{f12a}\\  
&& 
\hat{\Pi}_{ij} \boldsymbol{f}^{jk}=\boldsymbol{f}^{ik},  
\label{f12b}
\end{eqnarray}
where we suppressed the spatial arguments. 
Note, that no summation over repeating indexes is implied in Eqs. (\ref{f12a}),  (\ref{f12b}).   
For $\boldsymbol{f}^{12}$ Eq. (\ref{f12a})  implies 
\begin{equation}
 f^{12}_1=f^{12}_2, \, \,  \,  f^{12}_2=f^{12}_1. 
\end{equation}
 It follows that the function $\boldsymbol{f}^{12}({\bf X})$ can be parametrized by two independent functions $\alpha({\bf X})$ and $\beta({\bf X})$ 
\begin{equation}
\boldsymbol{f}^{12}=\alpha \left(
\begin{array}{c}
 1 \\ 1\\ 0 
\end{array}
\right) 
+\beta \left(
\begin{array}{c}
 0 \\ 0\\ 1 
\end{array}
\right) .
\label{f12_alphabeta_next}
\end{equation}
Applying the relations Eqs.  (\ref{f12b}) to Eq. (\ref{f12_alphabeta_next}), we obtain $\boldsymbol{f}^{34}=\boldsymbol{f}^{12}$, and 
\begin{eqnarray} 
&& 
 \boldsymbol{f}^{23}=\boldsymbol{f}^{14}=\alpha \left(
\begin{array}{c}
 0 \\ 1\\ 1 
\end{array}
\right) 
+\beta \left(
\begin{array}{c}
 1 \\ 0\\ 0 
\end{array}
\right),
\label{f23_alphabeta_2} \\ 
&& 
\boldsymbol{f}^{13}=\boldsymbol{f}^{24}=\alpha \left(
\begin{array}{c}
 1 \\ 0 \\ 1 
\end{array}
\right) 
+\beta \left(
\begin{array}{c}
 0 \\ 1 \\ 0 
\end{array}
\right) .
\label{f12_alphabeta-next2}
\end{eqnarray}
In terms of the functions $\alpha({\bf X})$ and $\beta({\bf X})$,  the general solution for the wave function acquires the form 
\begin{eqnarray}
\nonumber && 
 \left(\begin{array}{c}
        \chi_1({\bf X}) \\ \chi_2({\bf X}) \\ \chi_3({\bf X})
       \end{array}\right) =
\left(\begin{array}{c}
        \chi_1^0({\bf X}) \\ \chi_2^0 ({\bf X})\\ \chi_3^0({\bf X})
       \end{array}\right)  
+\left\{ \left(\int_{|{\bf r}'_1-{\bf r}'_2|=R_0}+\int_{|{\bf r}'_3-{\bf r}'_4|=R_0}\right) 
\left(\begin{array}{c} \alpha({\bf X}') \\ \alpha({\bf X}') \\  \beta({\bf X}') \end{array}\right) \right. \\
\nonumber && 
\left.
+\left(\int_{|{\bf r}'_1-{\bf r}'_3|=R_0}+\int_{|{\bf r}'_2-{\bf r}'_4|=R_0} \right) 
  \left(\begin{array}{c} \alpha({\bf X}') \\ \beta({\bf X}') \\  \alpha({\bf X}') \end{array}\right) 
+\left(\int_{|{\bf r}'_1-{\bf r}'_4|=R_0} +\int_{|{\bf r}'_2-{\bf r}'_3|=R_0}\right) 
 \left(\begin{array}{c} \beta({\bf X}') \\  \alpha({\bf X}') \\   \alpha({\bf X}') \end{array}\right) \right\}
G_E({\bf X}-{\bf X}') d{\bf X}',\\
\label{genSol-alpha-beta}
\end{eqnarray}
Now we apply the boundary condition Eq. (\ref{BP_chi}) to the general form Eq. (\ref{genSol-alpha-beta}), and derive equations for the functions $\alpha$ and $\beta$. For instance, the application of the boundary condition Eq. (\ref{BP_chi}) at $|{\bf r}_1-{\bf r}_2|=R_0$ for $\nu=2$ channel leads to equation 
\begin{eqnarray}
\nonumber && 
 \partial_{r_{12}}(\chi_1^0 + \chi_2^0)\bigg|_{|{\bf r}_{12}|=R_0} + 
2\partial_{r_{12}} \left\{\left[\int_{|{\bf r}'_{12}|=R_0} + \int_{|{\bf r}'_{34}|=R_0}\right] G_E({\bf X}-{\bf X}') \alpha({\bf X}')  d{\bf X}' + \right. \\ 
\nonumber && 
\left. 
\left[\int_{|{\bf r}'_{13}|=R_0} + \int_{|{\bf r}'_{14}|=R_0}+\int_{|{\bf r}'_{23}|=R_0}+\int_{|{\bf r}'_{24}|=R_0}\right] G_E({\bf X}_i-{\bf X}'_i) \left(\alpha({\bf X}')  +\beta({\bf X}') \right) d{\bf X}' \right\}\bigg|_{|{\bf r}_{12}|=R_0} = \\
\nonumber && 
-\frac{1}{a_2} \left\{(\chi_1^0 + \chi_2^0)\bigg|_{|r_{12}|=R_0} +
2 \left[\int_{|{\bf r}'_{12}|=R_0} + \int_{|{\bf r}'_{34}|=R_0}\right] G_E({\bf X}-{\bf X}') \alpha({\bf X}')   d{\bf X}' + \right. \\ 
\nonumber && 
\left. 
\left[\int_{|{\bf r}'_{13}|=R_0} + \int_{|{\bf r}'_{14}|=R_0}+\int_{|{\bf r}'_{23}|=R_0}+\int_{|{\bf r}'_{24}|=R_0}\right]
 G_E({\bf X}-{\bf X}') \left(\alpha({\bf X}')  +\beta({\bf X}') \right) d{\bf X}'
 \right\}\bigg|_{|{\bf r}_{12}|=R_0}. 
\end{eqnarray}
On the left hand side one can leave only the most singular term for $r_{12}\rightarrow R_0$, in which the the derivative of the Green function is taken by the variable $r_{12}$, normal to the scattering surface at which the boundary condition is imposed. We obtain  
\begin{eqnarray}
\nonumber && 
\int_{|{\bf r}'_{12}|=R_0}\partial_{r_{12}} G_E({\bf X}_i-{\bf X}'_i) \left(2 \alpha({\bf X}') \right) d{\bf X}' \bigg|_{|{\bf r}_{12}|=R_0}  = \mathcal{I}_0^{(2)}-\frac{1}{a_2} \left\{
 \left[\int_{|{\bf r}'_{12}|=R_0} + \int_{|{\bf r}'_{34}|=R_0}\right] G_E({\bf r}_i-{\bf r}'_i) (2\alpha({\bf X}'))   d{\bf X}' + \right. \\ 
 && 
\left. 
\left[\int_{|{\bf r}'_{13}|=R_0} + \int_{|{\bf r}'_{14}|=R_0}+\int_{|{\bf r}'_{23}|=R_0}+\int_{|{\bf r}'_{24}|=R_0}\right]
 G_E({\bf X}-{\bf X}') \left(\alpha({\bf X}')  +\beta({\bf X}') \right) d{\bf X}'
 \right\}\bigg|_{|{\bf r}_{12}|=R_0}. 
\label{BC2-alpha-beta}
\end{eqnarray}
Here 
\begin{equation}
\mathcal{I}_0^{(2)}=-\left(\frac{1}{a_2}+\partial_{r_{12}}\right)(\chi_1^0 + \chi_2^0)\bigg|_{|{\bf r}_{12}|=R_0}
\label{Source2}
\end{equation}
denotes the source field, describing the incoming wave in the $F=2$ channel. 
Analogously, in the $F=0$ channel, we obtain 
\begin{eqnarray}
\nonumber && 
\int_{|{\bf r}'_{12}|=R_0}\partial_{r_{12}} G_E({\bf X}-{\bf X}') \left(\frac{2}{3}\alpha({\bf X}')  +
\beta({\bf X}') \right) d{\bf X}'\bigg|_{|{\bf r}_{12}|=R_0}   =  \\
\nonumber && 
\mathcal{I}_0^{(0)} -\frac{1}{a_0} \left\{\left[\int_{|{\bf X}'_{12}|=R_0} + \int_{|{\bf X}'_{34}|=R_0}\right]G_E({\bf X}_i-{\bf X}'_i) \left(\frac{2}{3}\alpha({\bf X}')  +
\beta({\bf X}') \right) d{\bf X}' + \right. \\ 
 && 
\left. 
\left[\int_{|{\bf r}'_{13}|=R_0} + \int_{|{\bf r}'_{14}|=R_0}+\int_{|{\bf r}'_{23}|=R_0}+\int_{|{\bf r}'_{24}|=R_0}\right]
 G_E({\bf X}-{\bf X}') \left(\frac{4}{3}\alpha({\bf X}')  +\frac{1}{3}\beta({\bf X}') \right) d{\bf X}'
 \right\}\bigg|_{|{\bf r}_{12}|=R_0},  
\label{BC0-alpha-beta}
\end{eqnarray}
where 
\begin{equation}
\mathcal{I}_0^{(0)}=-\left(\frac{1}{a_2}+\partial_{r_{12}}\right)\left[\frac{1}{3}(\chi_1^0 + \chi_2^0) + \chi_3^0\right]\bigg|_{|{\bf r}_{12}|=R_0} 
\end{equation}
denotes the source field, describing the incoming wave in the $F=0$ channel. 
Due to the permutation symmetry, the boundary conditions at other pairs of points $|{\bf r}_{ij}|=R_0$  do not lead to new independent equations. 

\subsubsection{Transition to relative coordinates and separation of singularity}
The center of mass coordinate of four atoms is given by ${\bf R}=\frac{1}{4}({\bf r}_1+{\bf r}_2+{\bf r}_3+{\bf r}_4)$. 
For the following calculations we introduce the set of relative (Jacobi) coordinates: 
\begin{eqnarray}
{\bf z}=({\bf r}_3-{\bf r}_4), & {\bf y}=({\bf r}_1-{\bf r}_2), & {\bf x}=\frac{1}{\sqrt{2}}[({\bf r}_3+{\bf r}_4)-({\bf r}_1+{\bf r}_2)].  
\label{relative_coordinatessup}
\end{eqnarray}
The boundary condition at $|{\bf r}_{12}|=R_0$ now acquires the  form  $|{\bf y}|=R_0$. 
We go to the center of mass system by integrating Eqs. (\ref{BC2-alpha-beta}), ({\ref{BC0-alpha-beta}) over the center of mass coordinate ${\bf R}$. The resulting equations depend only on the relative coordinates given by Eq. (\ref{relative_coordinates}). 
The free four-particle Green function in relative coordinates satisfies the equation  
\begin{equation}
(\nabla^2_{\bf x}+\nabla^2_{\bf y}+\nabla^2_{\bf z}+E)G_{E}({\bf x-x'} , {\bf y-y'}, {\bf z-z'})=- 2 \delta({\bf x-x'})  \delta({\bf y-y'}) \delta({\bf z-z'}), 
\label{4PGF-eq}
\end{equation}
which is an equation for a Green function of a free particle in 6 dimensions. In the case of negative energy $E=-|E|<0$ the solution of Eq. (\ref{4PGF-eq}) can be written in the form  
\begin{equation}
 G_{ E  } (\mathbf Z) = \vert E \vert^2 G_0(\sqrt{\vert E \vert } |\mathbf Z|),
\end{equation}
where 
\begin{equation}
 G_0(\xi) =\frac{K_2( \xi)}{4\pi^3 \xi^2 }, 
\end{equation}
and ${\bf Z}=({\bf x}, {\bf y}, {\bf z})$ is the 6-dimensional vector of relative coordinates. $K_2(\xi)$ denotes the modified Bessel function. 
The Fourier transformed Green function is given by  
\begin{equation}
 G_{0}({\bf K})=\frac{2}{\left|{\bf K}\right|^2+1}, 
\end{equation}
where ${\bf K}=({\bf k}_{\bf x}, {\bf k}_{\bf y}, {\bf k}_{\bf z})$ is the 6-dimensional wave vector.

To  simplify the form of STM equations further, we introduce the source fields corresponding specifically to $S=0$ and $S=2$ channels as follows 
\begin{equation}
f_0({\bf Z})=2\pi R_0\left(\frac{2}{3}\alpha({\bf Z})+\beta({\bf Z})\right), \, \,  f_2({\bf Z})=4\pi R_0\alpha({\bf Z}). 
\label{def_f0f2}
\end{equation} 
The boundary  $|{\bf r}_{12}|=R_0$ transforms in the relative coordinates to $|{\bf y}|=R_0$.  
Eqs. (\ref{BC2-alpha-beta}),  (\ref{BC0-alpha-beta}) develop singularities at the surfaces $|{\bf r}'_{12}|=|{\bf r}_{12}|=R_0$, which in relative coordinates transforms to $|{\bf y}|=|{\bf y}'|=R_0$,  when  the collision surface coincides with the surface at which the boundary condition is imposed. 
The singularities are dealt  with by  subtraction and addition of  
$f_2({\bf x}, R_0, {\bf z})$ or $f_0({\bf x}, R_0, {\bf z})$ for Eqs. (\ref{BC2-alpha-beta}),  (\ref{BC0-alpha-beta}) respectively, similarly to the procedure described in Ref. \cite{Petrov}.
Finally, we introduce dimensionless coordinates by re-scaling $\mathbf x \to \mathbf x /\sqrt{\vert E \vert}$. Then the STM equations acquire the following explicit form 
\begin{eqnarray}
\nonumber &&  
\frac{f_0({\bf x}, {\bf z})}{\gamma_0}=\frac{1}{3}(\chi_1^0({\bf x}, 0, {\bf z}) + \chi_2^0({\bf x}, 0, {\bf z})) + \chi_3^0({\bf x}, 0, {\bf z}) +  \\ 
\nonumber && 
 \int[f_0({\bf x'}, {\bf z}'))-f_0({\bf x}, {\bf z}))] 
G_0({\bf x}-{\bf x'}, 0,  {\bf z}-{\bf z}') d^2{\bf x}' d^2{\bf z'} +  \\
\nonumber && 
\int G_0 \left(\sqrt{({\bf x}+{\bf x'})^2+{\bf y'}^2+{\bf z}^2}\right) f_0({\bf x'},  {\bf y'})  d{\bf x}' d{\bf y}'+ \\ 
 \nonumber && 
2\int \left[G_0\left(\sqrt{{\bf z}^2+{\bf x}^2+\sqrt{2} {\bf z}\cdot{\bf x'} +(\sqrt{2}{\bf x}-{\bf z})\cdot{\bf z'}+{\bf x'}^2+{\bf z'}^2}\right) + \right. \\
\nonumber && 
\left.   
G_0\left(\sqrt{{\bf z}^2+{\bf x}^2 - \sqrt{2} {\bf z}\cdot{\bf x'}  - (\sqrt{2}{\bf x}+{\bf z})\cdot{\bf z'}+{\bf x'}^2+{\bf z'}^2}\right) \right]  \left(\frac{1}{3}f_0({\bf x}', {\bf z'}) + \frac{5}{9}f_2({\bf x}',  {\bf z'})  \right) d{\bf x}' d{\bf z'},  \\
\label{0-channel_fin} 
\end{eqnarray}
\begin{eqnarray}
\nonumber &&  
\frac{f_2({\bf x}, {\bf z})}{\gamma_2}=(\chi_1^0({\bf x}, 0, {\bf z}) + \chi_2^0({\bf x}, 0, {\bf z}))  +  \\ 
\nonumber && 
 \int[f_2({\bf x'}, {\bf z}')-f_2({\bf x}, {\bf z})] 
G_0({\bf x}-{\bf x'}, 0, {\bf z}-{\bf z}') d^2{\bf x}' d^2{\bf z'} +  \\
\nonumber && 
\int G_0\left(\sqrt{({\bf x}+{\bf x'})^2+{\bf y'}^2+{\bf z}^2}\right) f_2({\bf x'},  {\bf y'})  d{\bf x}' d{\bf y}' + \\ 
 \nonumber && 
2\int G_0\left(\sqrt{{\bf z}^2+{\bf x}^2+\sqrt{2} {\bf z}\cdot{\bf x'} +(\sqrt{2}{\bf x}-{\bf z})\cdot{\bf z'}+{\bf x'}^2+{\bf z'}^2}\right) 
\left(f_0({\bf x}',  {\bf z'})  +\frac{1}{6}f_2({\bf x}', {\bf z'}) \right) d{\bf x}' d{\bf z'}  +\\
\nonumber && 
2 \int G_0\left(\sqrt{{\bf z}^2+{\bf x}^2 - \sqrt{2} {\bf z}\cdot{\bf x'}  - (\sqrt{2}{\bf x}+{\bf z})\cdot{\bf z'}+{\bf x'}^2+{\bf z'}^2}\right) 
\left(f_0({\bf x}',  {\bf z'})  +\frac{1}{6}f_2({\bf x}', {\bf z'}) \right) d{\bf x}' d{\bf z}'.  \\
\label{2-channel_fin}
\end{eqnarray}
where all distances are measured in units of $1/\sqrt{\vert E \vert}$. 
Here, all microscopic scattering parameters enter just in the form of two constants, $\gamma_0$ and $\gamma_2$, which are defined as follows 
\begin{equation}
\gamma_0=\frac{\pi}{\ln\left(R_0\sqrt{|E|} e^{a_0/R_0}\right) }, \, \, \gamma_2=\frac{\pi}{\ln\left(R_0\sqrt{|E|}e^{a_2/R_0}\right)}. 
\label{gamma_02}
\end{equation}
Fourier transform of Eqs. (\ref{0-channel_fin}), (\ref{2-channel_fin}) is performed according to the following Fourier representations of the Green functions 
\begin{eqnarray}
\nonumber && 
G_0\left(\sqrt{({\bf x}+{\bf x'})^2+{\bf y'}^2+{\bf z}^2}\right)= \\
&& 
\int\frac{d^2{\bf k}}{(2\pi)^2}\frac{d^2{\bf p}}{(2\pi)^2}\frac{d^2{\bf q}}{(2\pi)^2}\frac{2}{1+k^2+p^2+q^2} \exp\left[i\left\{{\bf k}\cdot {\bf z}+{\bf p}\cdot({\bf x}+{\bf x'})-{\bf q}\cdot{\bf y'}\right\}\right], 
\end{eqnarray}
and 
\begin{eqnarray}
\nonumber && 
G_0\left(\sqrt{{\bf z}^2+{\bf x}^2\pm\sqrt{2} {\bf z}\cdot{\bf x'}-({\bf z}\mp\sqrt{2}{\bf x})\cdot{\bf z'}+{\bf x'}^2+{\bf z'}^2}\right) = \\ 
&& \int\frac{d^2{\bf k}}{(2\pi)^2}\frac{d^2{\bf p}}{(2\pi)^2}\frac{d^2{\bf q}}{(2\pi)^2}\frac{2}{1+k^2+p^2+q^2} 
\exp\left[i\left\{{\bf k}\cdot\frac{{\bf z}-{\bf z'}}{\sqrt{2}}+{\bf p}\cdot \left(\frac{{\bf z}}{\sqrt{2}}\pm{\bf x'}\right)+{\bf q}\cdot \left(\frac{{\bf z'}}{\sqrt{2}}\pm {\bf x}\right)\right\}\right].
\end{eqnarray}
Below we demonstrate the representation of  two typical terms in Eqs. (\ref{0-channel_fin}), (\ref{2-channel_fin}) by Fourier components, from which the Fourier transform becomes obvious 
\begin{eqnarray}
\nonumber && 
\int G_0 \left(\sqrt{({\bf x}+{\bf x'})^2+{\bf y'}^2+{\bf z}^2}\right) f({\bf x'},  {\bf y'})  d{\bf x}' d{\bf y}'=\\
\nonumber && 
\int\frac{d^2{\bf k}}{(2\pi)^2}\frac{d^2{\bf p}}{(2\pi)^2}\frac{d^2{\bf q}}{(2\pi)^2}\frac{2}{1+k^2+p^2+q^2}
e^{i{\bf p}\cdot({\bf x}+{\bf x'})}e^{-i{\bf q}\cdot{\bf y'}}e^{i{\bf k}\cdot {\bf z}}f({\bf x'}, {\bf y'}) 
=\\ 
&& 
\int\frac{d^2{\bf k}}{(2\pi)^2}\frac{d^2{\bf p}}{(2\pi)^2}\frac{d^2{\bf q}}{(2\pi)^2}\frac{2}{1+k^2+p^2+q^2}f(-{\bf p},  {\bf q})
e^{i({\bf p}\cdot{\bf x}+{\bf k}\cdot {\bf z})}. 
\label{Fourier1}
\end{eqnarray}
This leads to the first term in the right hand side of Eq. (12) in the main text of the paper. 
Furthermore  
\begin{eqnarray}
\nonumber && 
\int G_0\left(\sqrt{{\bf z}^2+{\bf x}^2+\sqrt{2} {\bf z}\cdot{\bf x'} +(\sqrt{2}{\bf x}-{\bf z})\cdot{\bf z'}+{\bf x'}^2+{\bf z'}^2}\right) f({\bf x}', {\bf z'})  = \\ 
\nonumber && 
\int\frac{d^2{\bf k}}{(2\pi)^2}\frac{d^2{\bf p}}{(2\pi)^2}\frac{d^2{\bf q}}{(2\pi)^2}\frac{2}{1+k^2+p^2+q^2}
\int\frac{d^2{\bf k'}}{(2\pi)^2}\frac{d^2{\bf q'}}{(2\pi)^2}\int d{\bf x}' d{\bf y}' f({\bf k'}, {\bf q'}) 
e^{i{\bf x'}\cdot({\bf k'}+{\bf p})} e^{i{\bf z'}\cdot\left({\bf q'}+\frac{{\bf q}-{\bf k}}{\sqrt{2}}\right)}
e^{i{\bf z}\cdot\left(\frac{{\bf k}+{\bf p}}{\sqrt{2}}\right)} e^{i{\bf x}\cdot{\bf q}}=\\ 
&& 
\int\frac{d^2{\bf k}}{(2\pi)^2}\frac{d^2{\bf p}}{(2\pi)^2}\frac{d^2{\bf q}}{(2\pi)^2}\frac{2}{1+k^2+p^2+q^2} 
f\left(-{\bf p}, \frac{{\bf k}-{\bf q}}{\sqrt{2}}\right) e^{i\frac{{\bf k}+{\bf p}}{\sqrt{2}}\cdot{\bf z}} 
e^{i{\bf q}\cdot{\bf x}}.
\end{eqnarray}
Introducing new variables 
\begin{equation}
{\bf P}=-\frac{{\bf q}+{\bf p}}{\sqrt{2}}, \, \, \, {\bf Q}=\frac{{\bf q}-{\bf p}}{\sqrt{2}}, 
\label{PQ}
\end{equation}
we obtain 
\begin{equation}
-{\bf p}=\frac{1}{\sqrt{2}}({\bf P}+{\bf Q}), \, \, \, \frac{{\bf k}-{\bf q}}{\sqrt{2}}=\frac{1}{\sqrt{2}}({\bf k}+{\bf P}-{\bf Q}) , 
\label{pq}
\end{equation}
which leads to the last term in the right hand side of Eq. (12) in the main text of the paper with $s=-1$.

\end{document}